\newcommand{\eat}[1]{}
\newcommand{\stitle}[1]{\noindent{\bf #1}}
\documentclass[sigconf,nonacm]{acmart}
\usepackage{tcolorbox}
\usepackage{lipsum}
\usepackage{enumitem}
\usepackage[utf8]{inputenc}
\usepackage{mdframed}
\AtBeginDocument{%
  }

\setcopyright{acmlicensed}
\copyrightyear{2024}
\acmYear{2024}
\acmDOI{}
\acmConference[KDD Cup'24]{2024 KDD Cup Workshop for Retrieval Augmented Generation}
\acmISBN{}

\begin{document}

\title{Winning Solution For Meta KDD Cup' 24}

\author{Yikuan Xia}
\authornote{Both authors contributed equally to this research.}
\email{wfl00014@pku.edu.cn}
\affiliation{%
  \institution{Peking University}
  \city{Beijing}
  \country{China}
}

\author{Jiazun Chen}
\authornotemark[1]
\email{chenjiazun@stu.pku.edu.cn}
\affiliation{%
  \institution{Peking University}
  \city{Beijing}
  \country{China}
}

\author{Jun Gao}
\authornote{Corresponding Author}
\email{gaojun@pku.edu.cn}
\affiliation{%
  \institution{Peking University}
  \city{Beijing}
  \country{China}
}


\begin{abstract}
This paper describes the winning solutions of all tasks in Meta
KDD Cup ’24 from \textbf{db3} team. The challenge is
to build a RAG system from web sources and knowledge graphs. We
are given multiple sources for each query to help us answer the question. The CRAG challenge involves three tasks: (1) condensing information from web pages into accurate answers, (2) integrating structured data from mock knowledge graphs, and (3) selecting and integrating critical data from extensive web pages and APIs to reflect real-world retrieval challenges. Our solution for Task \#1 is a framework
of web or open-data retrieval and answering. The large language model (LLM) is tuned for better RAG performance and less hallucination.  Task \#2 and Task \#3 solutions are based on a regularized API set for domain questions and the API generation method using tuned LLM. Our knowledge graph API interface extracts directly relevant information to help LLMs answer correctly. Our solution achieves 1st place on all three tasks, achieving a score of 28.4\%, 42.7\%, and 47.8\%, respectively. 

\end{abstract}



\maketitle

\section{Introduction}

Ensuring the trustworthiness of language model (LLM) responses is critical due to the persistent issue of hallucination, where models generate inaccurate or ungrounded answers. Studies show that GPT-4's accuracy for fast-changing facts is often below 35\%~\cite{yang2024cragcomprehensiverag}. Retrieval-Augmented Generation (RAG) ~\cite{rag:conf/nips/2020,selfragconf/iclr/2024} offers a promising solution by integrating external information retrieval with LLMs to provide grounded answers. Despite its potential, RAG faces challenges in selecting relevant information, reducing latency, and synthesizing complex answers.

To address these issues, Meta introduces the Meta Comprehensive RAG Challenge (CRAG) as a 2024 KDDCup event, aiming to benchmark RAG systems with clear metrics and evaluation protocols to drive innovation and advance solutions in this field. CRAG Benchmark encompasses five domains, eight question types, varying answer timelines, and a range of entity popularity, including head, torso, and tail facts, as well as simple and complex question formats to test reasoning and synthesis capabilities. Each query has a time budget of 30 seconds, which also poses an efficiency challenge to the candidate solutions.

In detail, the CRAG challenge consists of three tasks:

\begin{enumerate}
    \item \textbf{Task \#1: Web-based Retrieval Summarization.} There are five web pages per question to identify and condense relevant information into accurate answers.
    \item \textbf{Task \#2: Knowledge Graph and Web Augmentation.} Mock APIs are provided to access structured data from mock knowledge graphs to integrate information into comprehensive answers.
    \item \textbf{Task \#3: End-to-End RAG.} 50 web pages and Mock APIs access per question are provided to select and integrate the most important data, reflecting real-world information retrieval challenges.
\end{enumerate}

The author's team, db3, participates in the contest and achieves first place in the three tasks, gaining a score of 28.4\%, 42.7\%, and 47.8\%, respectively. This paper describes the author's solution to the three tasks. The difference between the three challenges is that the information sources are different. Since Task \#2 and Task \#3 are provided with both sources, we describe the solution of these two tasks together. Our code is available on GitLab~\footnote{https://gitlab.aicrowd.com/jiazunchen/kdd2024cup-crag-db3}.

In the remainder of the paper, we discuss the web retrieval module and related model adjustment in Sec.~\ref{sec:1}, and we discuss the knowledge graph extraction module and related model adjustment in Sec.~\ref{sec:23}. We conclude our work and look into future works in Sec.~\ref{sec:conclusion}.

\section{Solution to Task \#1}
\label{sec:1}

In this section, we will propose our solution to Task \#1. Specifically, we will present the pipeline of processing web pages, which is also used in Task \#3. We will also describe the pipeline of tuning the base LLM, which may be similar for other purposes in Task \#2 and Task \#3.

\subsection{Framework for Solving Task \#1}

\begin{figure} [t]
    \centering
    \vspace{0.6cm}
    \includegraphics[width=1\columnwidth]{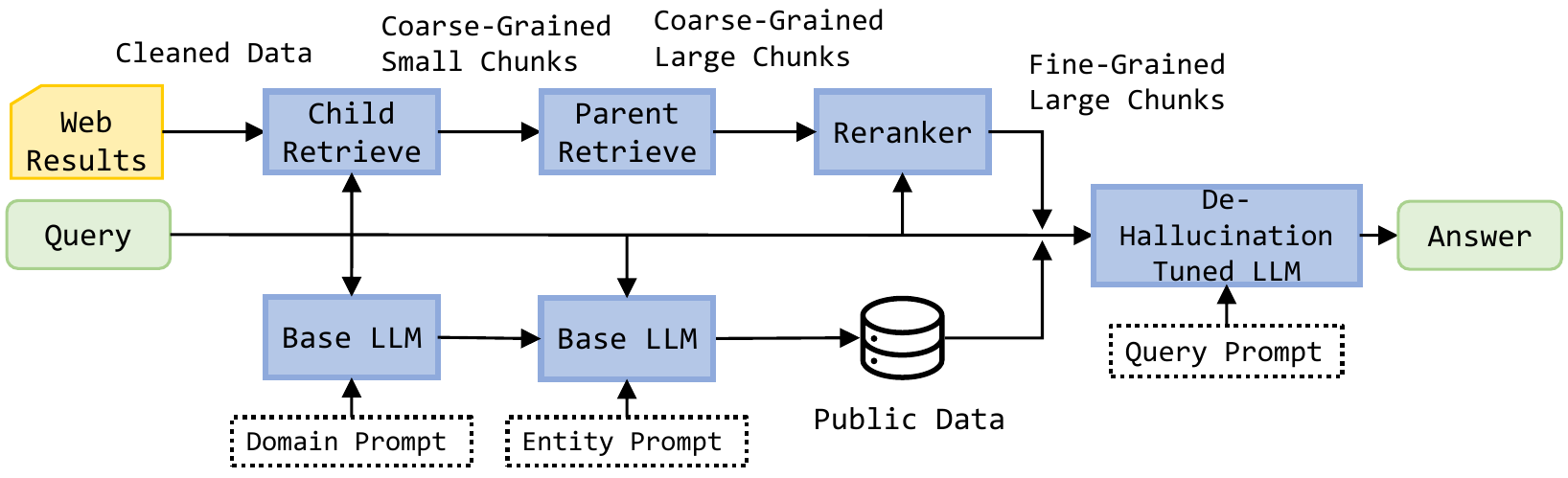}
    \caption{Illustration of Task 1 framework.}
    \label{fig:task1framework}
\end{figure}

Figure ~\ref{fig:task1framework} illustrates our framework for Task 1. We employ two pathways to retrieve information and ultimately use a tuned LLM to answer the questions, which relieves the hallucination problem.  Next, we will separately introduce each pathway and the adjustments made to the LLM.

\subsection{Web Retrieval Pathway}

We follow the widely used retriever and reranker framework for our web retrieval pathway.

\stitle{Text Extraction from Web Pages.} 
For the retrieval process, the text extraction module processes the HTML contents of search results, converting them into plain text chunks. Specifically, we use the BeautifulSoup library\footnote{https://www.crummy.com/software/BeautifulSoup/} to extract text content from raw HTML. We use the CharacterTextSplitter from LangChain\footnote{https://www.langchain.com/} to split the text into chunks.

\stitle{Parent-Child Chunk Retriever.} 
The Retriever ranks the text chunks by calculating the similarity between the query and each text chunk. Specifically, the retriever analyzes the keywords along with semantics within the query and matches them with the content of the text chunks. We refer to the open-source retriever leaderboards for retriever selection to ensure the most effective retriever is used. The final candidates are bge-base-en-v1.5~\cite{bge_embedding} and bce-embedding-base\_v1~\cite{youdao_bcembedding_2023}, producing similar results. The submitted version is based on bge-base-en-v1.5.

Since relatively smaller chunks have better retrieval precision and relatively larger chunks retain more information, we use the ParentDocumentRetriever from LangChain to manage a parent-child chunk split. We use the relatively small child chunks to retrieve from the question and the ParentDocumentRetriever to maintain the inclusion relationship. The smaller child chunks, such as individual sentences, are used for retrieval, while the larger parent chunks, which contain these retrieved child chunks, are typically whole paragraphs and are fed into the RAG system.

In the contest, larger parent chunk size will result in a larger context for LLM, leading to more inference time. So, to balance the time issue, we select parent\_chunk\_size=700, \\child\_chunk\_size=200 as a baseline. Under the submission constraints, we tune these hyperparameters, finding that \\parent\_chunk\_size=500,1000,2000 is also acceptable.

\stitle{Reranker.} The results from the retriever can be regarded as a coarse selection. In this challenge, we set the retriever to return $recall\_k$ text chunks. These chunks contain potentially relevant information, but not all are equally important or relevant.

To more effectively utilize this information within the limited context, we introduce a reranker model. The reranker model performs a secondary screening by more finely evaluating and ranking the $recall\_k$ data chunks returned by the Retriever. Through this detailed screening process, we can identify the most valuable and relevant chunks, ultimately selecting the top $reranker\_k$. These top $reranker\_k$ chunks will serve as the basis for further processing and answering the questions, ensuring that the our answers are more accurate and reliable. 

We refer to the open-source reranker leaderboards for reranker selection. The final candidates are bge-reranker-v2-m3~\cite{li2023making,chen2024bge} and bce-reranker-base\_v1~\cite{youdao_bcembedding_2023}. The results are rather similar. The submitted version is based on bge-reranker-v2-m3.

To fit in the LLM context and given a limited running time, we set the recall number for the retriever $recall\_k=50$. For the number of chunks we fed to the LLM $reranker\_k$, we set it according to the parent\_chunk\_size. For example, we take $reranker\_k=5$ if parent\_chunk\_size=2000, and we would take $reranker\_k=10$ if parent\_chunk\_size=1000.

\subsection{Public Data Pathway}
A regular web retriever usually suffers greatly from insufficient information and misinformation. For some of the stable facts, we can gather public data to provide additional information. This information is preprocessed into a paragraph, which is combined with the content of the web search to help the LLM answer the question.

This information for different domains is constructed differently. For the movie domain, we preprocess the Oscar award information\footnote{https://www.kaggle.com/datasets/unanimad/the-oscar-award} and the Full MovieLens\footnote{https://www.kaggle.com/datasets/rounakbanik/the-movies-dataset?select=movies\_metadata.csv}. For finance, we preprocess the current pe-ratio, market cap and eps stats for every current stock in America. For music, we preprocess the Grammy award information\footnote{https://www.kaggle.com/datasets/unanimad/grammy-awards}.  Due to the limited information in other domains, no preprocessing is conducted.  Specifically, we serialize the entity's table/json data into natural language format using the structure "The [entity attribute] is [value of the entity attribute]." Specifically, we serialize the entity's table/json data into natural language format using the structure "The [entity attribute] is [value of the entity attribute]." For example, for a specific movie, the preprocessed format is: \textit{The title is "Rain Man." The director is Barry Levinson. The lead actors are Dustin Hoffman and Tom Cruise. The release year is 1988...}  

We store the preprocessed information using the corresponding entities as keys and then utilize LLM to locate the query entities through in-context learning.  Specifically, we first locate the problem domain based on \textbf{Domain Prompt}:
\vspace{0.3cm}\begin{mdframed}[linecolor=black, backgroundcolor=white, linewidth=1pt, roundcorner=0pt]
[\{"role": "system", "content": "You are an assistant expert in movie, sports, finance and music fields."\},\\
\{"role": "user", "content": "Please judge which category the query belongs to, without answering the query. You can only and must output one word in (movie, sports, finance, music). If the question doesn't belong to movie, sports, finance, music, please answer other. \\
Question: \{query\_str\}\\
Answer:
"\}]. 
\end{mdframed} \vspace{0.3cm} Because open domain questions often include content from other domains, we use "other" instead of "open".  Next, we design different prompts for various domains to query the entities. For example, for movie \textbf{Entity Prompt}:

\vspace{0.3cm}\begin{mdframed}[linecolor=black, backgroundcolor=white, linewidth=1pt, roundcorner=0pt]
[\{"role": "system", "content": "You are a helpful and honest assistant. Please, respond concisely and truthfully in \{token\_limit\} words or less. If you are not sure about the query, answer I don't know. There is no need to explain the reasoning behind your answers. "\},\\
\{"role": "user", "content": Given a query about movies, return the title of each movie in below formats.  \\
If multiple movie names are involved, connect with '\&\&'.\\
\#Examples:\\
Question:  which movie was created first, a walk to remember or the notebook?\\
Answer:    a walk to remember \&\& the notebook\\
......\\
\#Query:\\
Question: \{query\_str\}\\
Answer:
"\}].
\end{mdframed} \vspace{0.3cm}
Regarding how the entities in LLM's responses are linked to entities in the dataset, the required level of matching varies based on the characteristics of the question domain. Matching criteria range from strict to lenient, including exact character matches, substring inclusion, or similarity comparisons using embedding models.

\subsection{LLM Inference Module}
\label{sec:llm1}
According to the contest rules, the correct answer will be awarded 1 point, and the incorrect answer will be penalized for 1 point. So, a significant challenge during the contest is reducing hallucinations and dealing with invalid questions. We will present our LLM inference module to the two challenges.

\stitle{Base model.} We follow the contest instructions to use the LLama series LLM~\footnote{https://llama.meta.com/}. Considering the limited running time, we use the Llama-3-8B-instruct model as the base model.

\stitle{Basic Query Prompt.} We use the following basic prompt $p_{basic}$ to generate answers:
\vspace{0.3cm}\begin{mdframed}[linecolor=black, backgroundcolor=white, linewidth=1pt, roundcorner=0pt]

[\{"role": "system", "content": "You are a helpful and honest assistant. Please, respond concisely and truthfully in \{token\_limit\} words or less. Now is \{query\_time\}"\},\\
\{"role": "user","content": "Context information is below.\\
\{context\_str\}\\
Given the context information and using your prior knowledge, please provide your answer in concise style. End your answer with a period. Answer the question in one line only.\\
Question: \{query\_str\}\\
Answer:
"\}]

\end{mdframed} \vspace{0.3cm}
where token\_limit is the limit to the answer that we want to control (as an answer longer than 75 tokens will be truncated.), query\_time is the time when the question is asked, which is crucial in real-time questions, context\_str is formed by combining data from public retrieval and $reranker_k$ web retrievals using <doc> tokens, then truncated based on a maximum token limit of $4000$  (note that the previous retriever's chunk size is based on characters, while here it is based on tokens),  query\_str is the query.

\stitle{Reduce Hallucination using Prompt Control.} Hallucination can partly be controlled by prompt design. Certain instructions in the prompt can hold the LLM from generating wrong facts. E.g., we can use the following controlled prompt $p_{ctrl}$ to generate higher-quality answers:

\vspace{0.3cm}\begin{mdframed}[linecolor=black, backgroundcolor=white, linewidth=1pt, roundcorner=0pt]

[\{"role": "system", "content": "You are a helpful and honest assistant. Please, respond concisely and truthfully in \{token\_limit\} words or less. Now is \{query\_time\}"\},\\
\{"role": "user","content": "Context information is below.\\
\{context\_str\}\\
Given the context information and using your prior knowledge, please provide your answer in concise style. Answer the question in one line only. \\
If the question is based on false prepositions or assumptions, output "invalid question". For example, What's the name of Taylor Swift's rap album before she transitioned to pop? (Taylor Swift didn't release any rap album.)\\
If you are not sure about the question, output "i don't know"\\
Question: \{query\_str\}\\
Answer:
"\}]

\end{mdframed} \vspace{0.3cm}

Using $p_{ctrl}$ to control the generation can lead to better results, as for questions hard for the LLM to answer, it will probably avoid the penalty. For some invalid questions, the tuned model may point out that the questions are based on false prepositions. However, the overall results are not satisfying. So, we don't directly use this method to control hallucination.

\stitle{Reduce Hallucination from Fine-tuning.} Usually, through sufficient supervised fine-tuning (SFT), it's possible to make the LLM perform better on a particular task. So, we try fine-tuning the base model to reduce hallucination further in the contest.  

Through experiment, we find out that using $p_{ctrl}$ to generate answers will reduce hallucination, while in the meantime, some questions which can be answered correctly using $p_{basic}$ will be answered wrongly with "i don't know". So, we hope we can leverage the whole potential of the RAG system while hindering most of the wrong answers.

It's clear that for some continuing changing facts, it's impossible for the LLM to answer correctly if not provided with the fact in the context\_str. Our intuition is that we hope the LLM can answer correctly for the facts contained in the context\_str. For the facts that are not contained in the context\_str but in the LLM's internal knowledge, we hope the LLM can answer correctly, too. For the other queries, which are out of the RAG's capabilities, we hope the LLM can answer "i don't know" honestly. As common sense, such patterns may be learned by the LLM. For example, for the continuing changing finance problems, the answer accuracy is close to 0, so the LLM may learn to answer honestly "i don't know" for such questions.

So, we follow the following steps to generate the labels for SFT to meet our intuition:
\begin{itemize}
    \item 1. For the queries identified in the ground truth as invalid questions, we set the labels to "invalid question", hoping that the model can possess the capability to find questions based on false premises.
    \item 2. We generate answers using the $p\_{basic}$ prompt for each query in the training split to leverage the potential of our RAG system fully. We use prompt $p_{check\_gt}$ and LLM to judge whether the answer is correct. (The LLM here can be Llama3, or, more accurately, GPT4.\footnote{For long answers generated by Llama3, it appears that judging also by Llama3 would be inaccurate. The answer may be more favored by its generated model. As for short answers, this problem is relieved greatly.})
    \item 3.1 For the queries labeled as correctly answered, we label the query with the ground truth answer because the query is within the potential of the current RAG system.
    \item 3.2 For the queries labeled as wrongly answered, we use LLM to check whether the ground truth answer can be inferred from the context\_str using $p\_{context}$. (The LLM here can be Llama3 or, more accurately, GPT4.)
    \item 3.2.1 For the queries with context\_str that the LLM regards as irrelevant to the ground truth answer, we label these queries with "i don't know", as these questions can hardly be answered correctly.
    \item 3.2.2 For the queries with context\_str that the LLM regards as relevant to the ground truth answer, we label these queries with the ground truth answer. We hope the fine-tuned LLM can have a stronger comprehension ability from the noisy context containing hints of the correct answer.
\end{itemize}

We move the fine-tuning part after constructing the training data as above. Considering limited computation resources, we use LoRA~\cite{hu2021LoRA} to fine-tune the base model. Using LoRA models has another advantage. We can fine-tune several LoRA models each responsible for a subtask. Since LoRA parameters are tiny compared with the LLM parameters, it's easy to switch between the subtasks, which is time-saving. We tune the basic model for 2-3 epochs on the training set. The tuning hyperparameters can be found in the appendix. After tuning, there are three significant improvements:

\begin{enumerate}[label=\arabic*.]
    \item The answer style becomes closer to the short and direct answer as the ground truth answers, making it harder for the LLM to generate hallucinations during reasoning.
    \item The LLM can judge the "i don't know" case more accurately, leading to a better result.
    \item The LLM can deal with some false premise cases, saving many points.
\end{enumerate}

\stitle{Inference Acceleration.} We use vLLM to accelerate the inference process~\footnote{https://github.com/vllm-project/vllm}. However, though the latest vLLM library supports the LoRA framework, the graphics card driver in the test environment has some compatibility issues. So, we have to load each LoRA model completely when we aim to switch between multiple LoRA models. Each query can share the switching time through batch inference to meet the time requirement. Our submission contains versions that use and do not use vLLM. If the compatibility issue is solved, introducing vLLM can save much time.

\section{Solution to Task \#2 and Task \#3}

In this section, we will first introduce the main framework of our solution for Task \#2 and Task \#3. Then, we will propose our knowledge graph retrieval module based on a set of regularized APIs and API generation using a tuned LLM. The web retrieval and answer generation part of Task \#2 and Task \#3 is similar to the ones in solution to Task \#1, which will be presented briefly in this section.
\label{sec:23}
\subsection{Framework for Solving Task \#2 and Task \#3}

\begin{figure*} [t]
    \centering
    \includegraphics[width=2\columnwidth]{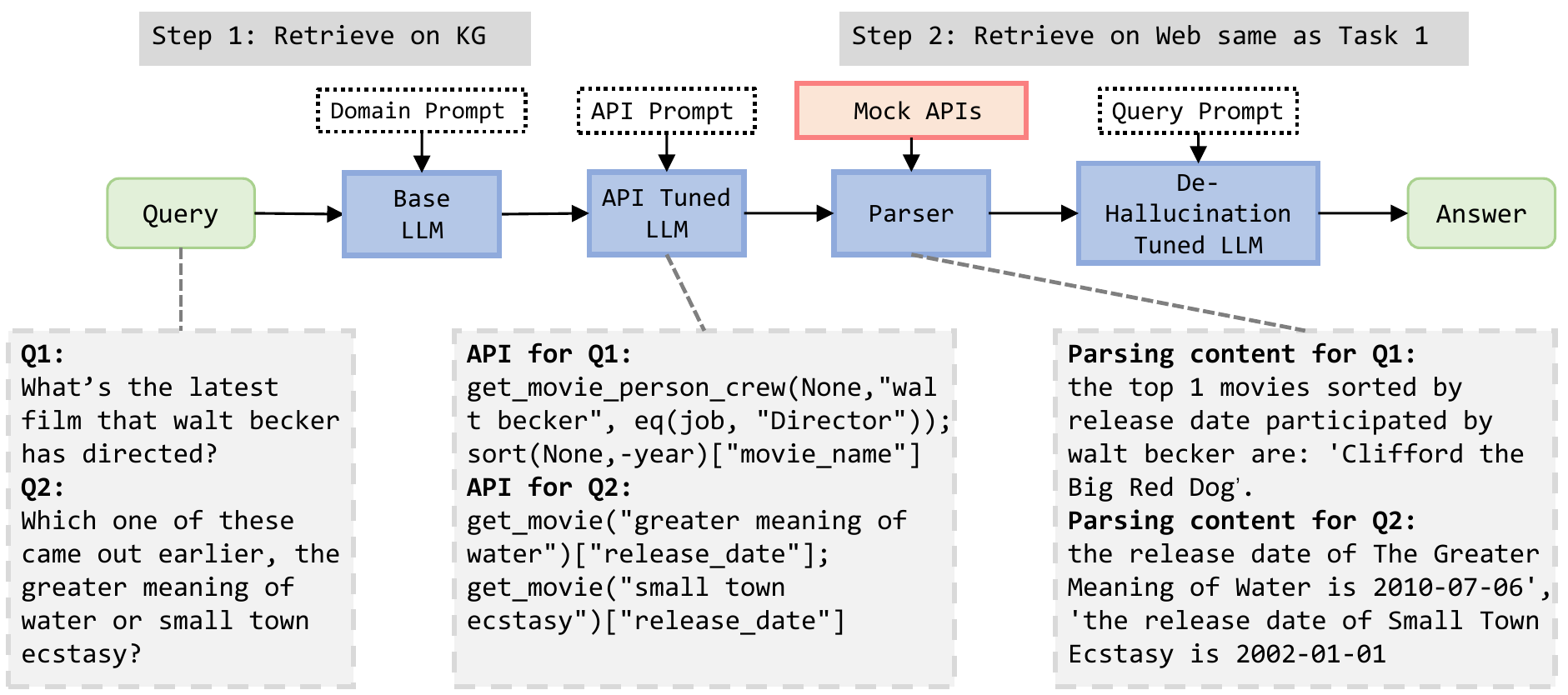}
    \caption{Illustration of Task \#2 and \#3 framework.}
    \label{fig:task2framework}
\end{figure*}
We observe that the information extracted from web pages is noisier than the information from the Mock APIs. Therefore, our framework separates the answer from the Mock API and web pages, and we prioritize the results based on the Mock API results. Once the results based on the Mock API content are not "i don't know", we accept the result and output directly. The framework of Task \#2 and \#3  is illustrated in Fig.~\ref{fig:task2framework}.

Notably, the web retrieval part for Task \#3 is slightly different from those in Task \#1 and Task \#2. Since Task \#3 has 50 web pages, the time budget should be taken into consideration. In the contest, a page snippet is provided for each web page. We use the reranker to find the top five related web pages based on the snippets, which makes the overall running time for the retriever acceptable, and then process them according to the method used in Task 1. Additionally, we don't use public data in Task \#2 and \#3 because it's basically covered by Mock APIs.

\subsection{Knowledge Graph Retrieval Module}

A Mock API knowledge graph is provided in Task \#2 and Task \#3. We will describe how we extract useful information from this API or knowledge graph. The main idea of our knowledge graph retrieval module is to use LLMs to generate a series of API calls, which extract the specific information the query is asking about.

\stitle{Regularization of the API}. Though the APIs provided contain rich information, it's quite hard to locate the exact information we want. The exact location of the information can be found by executing the code generated by the LLM. However, the code generation capability of a local 7B LLM is rather limited. So we design a regularized version of API. For each query, only one (or several) APIs are generated.  From the generated regularized APIs, we use a parsing system to get the generated results. The results are then converted to natural language to form the output of our retrieval module.

We take the movie domain as an example. The original movie API consists of the following API calls:
\begin{itemize}
    \item get\_person\_info (person\_name)$\rightarrow$ name (string): name of person; acted\_movies (list); directed\_movies (list); birthday (string); oscar\_awards
    \item get\_movie\_info (person\_name)$\rightarrow$ title (string); release\_date (string); original\_title (string); original\_language (string); budget (int); revenue (int); rating (float); genres (list); oscar\_awards; cast (list); crew(list)
    \item get\_year\_info (year)$\rightarrow$ movie\_list(list); oscar\_awards(list)
\end{itemize}

This actually forms a relational database behind the scenes. There are two relational tables: the PERSON table and the MOVIE table. For instance, the PERSON table has the columns: name, birthday, and the MOVIE table has the columns: title, release\_date, original\_title, original\_language, budget, revenue, rating, genres, year. There are three additional tables that have foreign keys referring to the PERSON table and MOVIE table, which records the relationship between person and movies: the CAST table, the CREW table and the OSCAR table. Each entity of the five tables can be constructed using the API calls. So theoretically, we can use SQL language to query information from this relational database, and converting the query to SQL language is a typical text2sql task~\cite{yu2018spider}, which is well studied. However, it's impossible to call the API multiple times to form a relational database and execute SQL on it due to time budget. So, we design a new type of regularized API that is easy to parse and execute and leverages the characteristics of relational databases.

For the movie domain, we set the following API:
\begin{itemize}
    \item get\_person (person\_name,condition)[key\_name], which is equal to \textit{SELECT key\_name FROM PERSON WHERE condition and name=person\_name}
    \item get\_movie(movie\_name,condition)[key\_name], which is equal to \textit{SELECT key\_name FROM MOVIE WHERE condition and title=movie\_name}
    \item get\_movie\_person\_X(movie\_name,person\_name,condition) [key\_name], (here X=CREW, CAST or OSCAR), which is equal to \textit{SELECT key\_name FROM MOVIE, PERSON, X WHERE condition and title=movie\_name, name=person\_name.}
\end{itemize}

Using these three regularized APIs, most of the information the query wants in this contest can be extracted with only a few API calls. These APIs can also be easily evaluated by the original APIs through a parsing system. More importantly, these APIs are easy to generate for the LLM, as essentially, only template selection and extraction of entity names have to be done by the LLM. There are no complex codes or SQL generation involved, which may lead to better performance. 

We design different APIs for the five different domains, but the design choices are all similar. The details of these APIs can be found in our released codes.

\stitle{Details of the API System.} How to make the conditions work in the above system is a question. Answering some of the questions in the contest involves a slight modification of the API system above. For instance, querying the latest Spielberg film actually requires the sort function. For some queries in the post-processing questions, numeric computations are involved. To meet these requirements, we have some other words for generation in our designed corpus:

\begin{itemize}
    \item We can use cmp (key\_name,value\_name) to set a condition. The cmp here can be neq, eq, ge and le, which represents not equal, equal, greater, less respectively. e.g., eq(gender,male), which means the condition of gender to be male. The condition can be a list of multiple conditions\\, e.g., [eq(gender,"male"),eq(character,"batman")].
    \item We can use ["len"] to get the output lengths of a result, and we can use AVG to get the average numeric value of the output.
    \item We can use sort(condition,sort\_key\_name) to get a sorted list which satisfies such condition, and the list is sorted using the sort\_key\_name. If we want descending sort, we can use -sort\_key\_name.
\end{itemize}

These are all simplified operators for coded functions or functions in SQL. Other functions include using * to represent the output of the last query and using * in another query, which is similar to the sublist function in SQL. This can perfectly fit the multi-hop query scenario. It's a pity that we haven't finished this part due to limited time in this contest, and we look forward to developing such functions in future systems.

\stitle{Parser.} We manually program the parser to meet the requirements of most queries in the development set. This may affect the generalization capability of this system, so using the standard SQL execution engine may be a better choice if the relational databases are presented. The output of the parser is converted to natural language so that even if irrelevant information is extracted, the LLM may be aware of that and refuse to answer.

\stitle{Examples of New Regularized APIs.} Here are three ideal examples of the new regularized API in the movie's domain:
\begin{itemize}
    \item Which one of these came out earlier, \textit{the greater meaning of water} or \textit{small town ecstasy}? $\rightarrow$ get\_movie("\textit{greater meaning of water}")["release\_date"]; get\_movie("\textit{small town ecstasy}")["release\_date"]
    \item Who won the best actor oscar for their performance in a movie in 2012? $\rightarrow$get\_movie\_person\_oscar\\
    (None,None,[eq(year,2012),eq(category,"best actor")\\
    ,eq(winner,"true")])["name"]
    \item What's the latest film that \textit{walt becker} has directed? $\rightarrow$ get\_movie\_person\_crew(None,"\textit{walt becker}", eq(job, "Director")); sort(None,-year)["movie\_name"]
\end{itemize}
An example of the full pipeline can also be found in Fig.~\ref{fig:task2framework}.

\stitle{API Generation.} After setting down the rules and parser of the new regularized API system, we move to solving the API generation problem. We use prompts to help the LLM generate as many valid APIs as possible to help us extract information. Neglecting the system message, the API generation prompt $p_{gen\_API}$ is as follows:
\vspace{0.3cm}\begin{mdframed}[linecolor=black, backgroundcolor=white, linewidth=1pt, roundcorner=0pt]
You are given a query about movies, and several APIs to get information from a database How to collect useful information from the database using the given APIs.\\
The schema of entities is as follows:\\
\{Schema\_info\}\\
The API rules are below:\\
\{API\_rules\}\\
Here are some examples:\\
\{ICL\_examples\}\\
Generate the answer only using the information from the query. Please strictly follow the format in the examples and APIs, you do not have to provide the code, only the use of API in the examples. The only allowed format is multiple lines of get\_X,sort. (sort is optional) Please complete the answer only:\\
Query:\{query\_str\}\\
Answer:\\ 
\end{mdframed} \vspace{0.3cm}
where Schema\_info are some descriptions about the underlying relational database schema, restricting the LLM to generate key names in the relational tables, API\_rules are the rules for generating the API described in the above subsection, query\_str is the question, and ICL\_examples are some in-context learning selected examples of query and API pairs. 

Here, the in-context learning examples are selected manually iteratively. First a few examples are selected, and we generate 100 examples using the LLM. Then the queries with the wrongly generated examples are added to the examples. Notably, selecting relevant ICL examples may be extremely effective in this scenario, as for similar queries, only the entity names have to be substituted. We haven't realized this function, but we believe it may have great results. 
Details of this prompt for different domains can be found in our source code. Here, we present some details of the prompt for movies in the Appendix.

\stitle{Fine-tuning for API Generation.} Through experiment, we observe that using $p_{API\_gen}$ still requires the LLM to have relatively strong capabilities. Strong LLMs, e.g., GPT-4, perform better than the local Llama3 8B. So we hope fine-tuning the local LLM can help boost the performance of API Generation. For the fine-tuning ground truth data, We use GPT-4 and $p_{API\_gen}$ to generate a first version of ground truth APIs for convenience. Then, we manually label the ground truth APIs for higher quality. We also use LoRA to fine-tune our base Llama3 model, as we need to efficiently switch between different LoRA parameters under the time budget. We also fine-tune the base model for 2-3 epochs, and the hyperparameters are also listed in the Appendix.

\subsection{LLM Inference Module}
\label{sec:2}
We follow the same framework in Sec.~\ref{sec:1} to fine-tune the model for LLM inference. The tuned model has a higher priority than the model tuned on web page results.

\section{Conclusion}
The Meta KDDCup 24 competition is a unique challenge due to the various types of information sources and the changing facts, which are difficult for the LLMs. We have presented how we addressed these challenges successfully in all three tasks of the contest. Our solution for Task \#1 is a framework
of web or open-data retrieval and tuned LLM for question answering. The solution to Task \#2 and Task \#3 is based on a regularized API set for domain questions and the API generation method using tuned LLM. We will further look into the balance of efficiency and effectiveness in RAG and a refined API parsing system for RAG to extract information from structured sources in the future.
\label{sec:conclusion}

\section{Acknowledgement}

This work was supported by the National Natural Science Foundation of China (NSFC) under Grant No. 62272008.

\bibliographystyle{ACM-Reference-Format}
\bibliography{sample-base}

\appendix

\section{Detail Evaluation of Our Solution}
\begin{figure*}[t]
    \centering
    \includegraphics[width=0.8\textwidth]{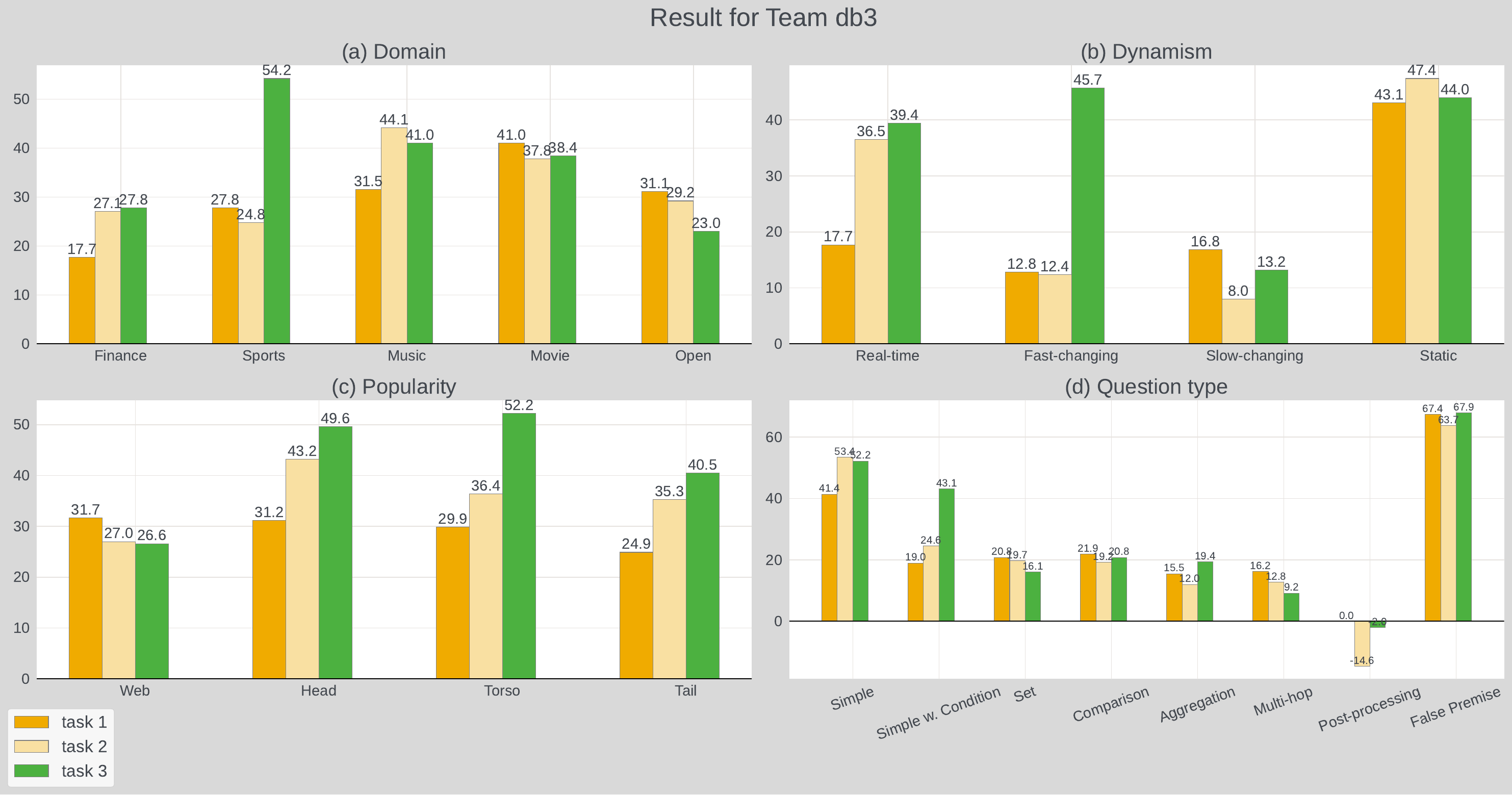}
    \caption{Detail Evaluation of Team \textbf{db3} Solution}
    \label{fig:yourlabel}
\end{figure*}

The organizer presented the detail evaluation of our solution, and we list it in Fig.~\ref{fig:yourlabel}. As we can see extracting useful information from KGs can significantly boost our performance on real-time and fast-changing queries. Accuracy boost on the finance domain is also sound, even outperforming some commercial RAG systems~\cite{yang2024cragcomprehensiverag}.

\section{LoRA and FineTuning Hyperparameters.}

The LoRA and Finetuning hyperparameters used in tuning the base model and API generation module is listed in Tab.~\ref{tab:my-table}:

\begin{table}[ht]
\caption{LoRA and FineTuning Hyperparameters.}

\begin{tabular}{ll}
\hline
Name                            & Value                                                                                                                    \\ \hline
LoRA\_alpha                     & 16                                                                                                                       \\
LoRA\_dropout                   & 0.1                                                                                                                      \\
LoRA\_r                         & 8                                                                                                                        \\
target\_modules                 & \begin{tabular}[c]{@{}l@{}}{[}"k\_proj", "q\_proj", "v\_proj",\\  "up\_proj", "down\_proj", "gate\_proj"{]}\end{tabular} \\
bias                            & "none"                                                                                                                   \\
4-bit                           & True                                                                                                                     \\
max\_seq\_length                & 2048/4096                                                                                                                \\
per\_device\_train\_batch\_size & 1                                                                                                                        \\
gradient\_accumulation\_steps   & 4                                                                                                                        \\
optim                           & "adamw\_hf"                                                                                                              \\
learning\_rate                  & 2e-4                                                                                                                     \\
max\_grad\_norm                 & 0.3                                                                                                                      \\
scheduler                       & "cosine", warm\_up\_ratio=0.1                                                                                            \\ \hline
\end{tabular}
 
\label{tab:my-table}
\end{table}

\section{Used Prompts.}
\eat{
\subsection{Prompt for domain identification.}
Neglecting the system message, $p_{domain}$ is as follows:\

\vspace{0.3cm}\begin{mdframed}[linecolor=black, backgroundcolor=white, linewidth=1pt, roundcorner=0pt]

[
        {"role": "system", "content": f"You are an assistant expert in movie, sports, finance and music fields."},\\
        {"role": "user", "content": "Please judge which category the query belongs to, without answering the query. you can only and must output one word in (movie, sports, finance, music) If the question doesn't belong to movie, sports,finance, music, please answer other. \\
        Query: \{query\_str\} \\
        Category:'},
]

\end{mdframed} \vspace{0.3cm}

}
\subsection{Prompt for checking the correctness of an answer.} 
Neglecting the system message, $p_{check\_gt}$ is as follows:

\vspace{0.3cm}\begin{mdframed}[linecolor=black, backgroundcolor=white, linewidth=1pt, roundcorner=0pt]

INSTRUCTIONS =\\
\# Task: \\
You are given a Question, a model Prediction, and a list of Ground Truth answers, judge whether the model Prediction matches any answer from the list of Ground Truth answers. Follow the instructions step by step to make a judgement.\\
1. If the model prediction matches any provided answers from the Ground Truth Answer list, "Accuracy" should be "True"; otherwise, "Accuracy" should be "False".\\
2. If the model prediction says that it couldn't answer the question or it doesn't have enough information, "Accuracy" should always be "False".\\
3. If the Ground Truth is "invalid question", "Accuracy" is "True" only if the model prediction is exactly "invalid question".\\
\# Output:\\
Respond with only a single JSON string with an "Accuracy" field which is "True" or "False".\\
\# Examples:\\
\{ICL\_examples\}\\
\# Query:\\
Question: \{query\_str\}\\
Ground truth: \{gt\_str\}\\
Prediction: \{our\_str\}\\
Accuracy:
 
\end{mdframed} \vspace{0.3cm}
where ICL\_examples are some in-context learning examples for the query, gt\_str is the ground truth answer, and our\_str is our generated answer.

\subsection{Prompt for Checking the RAG Context.} 
Neglecting the system message, $p_{context}$ is as follows:

\vspace{0.3cm}\begin{mdframed}[linecolor=black, backgroundcolor=white, linewidth=1pt, roundcorner=0pt]

We have the following context information:\\
\{context\_str\}\\
We have a question:  \{query\_str\}\\
The ground truth answer is: \{gt\_str\}\\
Is the ground truth answer mentioned in the context information? Answer with yes or no.
 
\end{mdframed} \vspace{0.3cm}
where query\_str is the question, gt\_str is the ground truth answer, and context\_str is the context from our retrieval system.

\subsection{Prompt for API Gereration (Movie)}

Neglecting the system message, $p_{API\_gen}$ for the movie domain is as follows:

\vspace{0.3cm}\begin{mdframed}[linecolor=black, backgroundcolor=white, linewidth=1pt, roundcorner=0pt]
You are given a query about movies, and several APIs to get information from a database How to collect useful information from the database using the given APIs.\\
The schema of entities are as follows:\\
Movie:- title (string): title of movie\\
...\\
- year (string): year of the movie\\
Person:- name (string): name of person\\
- birthday (string): string of person's birthday, in the format of "YYYY-MM-DD"\\
Besides we have the concat tables for the concat of these two basic entities:\\
Cast Movie Person: list of cast members of the movie and their roles. The schema of the cast member entity is:\\
-'movie\_name':name of the movie,\\
...\\
-'year'(string):the year of casting\\
Crew Movie Person: list of crew members of the movie and their roles.\\
-'movie\_name':name of the movie,\\
...\\
-'year'(string):the year of crewing\\
Oscar info: list of oscar awards, win or nominated, in which the movie was the entity. The schema for oscar award entity are:\\
'year' (int): year of the oscar ceremony,\\
...\\
'winner' (bool): whether the person won the award\\
\\
The API rules are below:\\
1.you can use cmp(key\_name,value\_name) to set a condition, the cmp here can be neq,eq,ge,le, which represents not equal,equal, greater, lesser respectively. e.g eq(gender,male), which means the contion of gender to be male, ge(revenue,10), which means the condition of revenue greater than 10. the condition can be a list of multiple conditions,\\
e.g. [eq(gender,"male"),eq(character,"batman")] you can add condition to the last parameter of get\_X\_info(X\_key\_value,\\
condition)\\
2.you can use get\_movie(movie\_name,condition)[key\_name] to search movie\_name for the most relevant result under such condition and query the key\_name attribute of it. the key names valid to use with get\_movie\_info is the key of the movie entities.\\
...\\
8.By default we output the first element of one list, however if you want it all, you can add ALL in the front of the command, e.g. ALL, ALL get\_movie\_person\_crew("batman","Jack",1997), represent get ALL Jack crews of batman movies in 1997. You can use [:n] to represent take the first n result of the list\\

Here are some examples:\\
\{ICL\_examples\}\\
Generate the answer only using the information from the query. Please strictly follow the format in the examples and APIs, you do not have to provide the code, only the use of API in the examples. The only allowed format is multiple lines of get\_X,sort. (sort is optional) Please complete the answer only:\\
Query:\{query\_str\}\\
Answer:\\ 
\end{mdframed} \vspace{0.3cm}

\end{document}